\documentclass[manuscript]{aastex}

\shorttitle{ChromaStarDB}
\shortauthors{Short}

\begin{document}


\title{ChromaStarDB: SQL database-driven spectrum synthesis, and more}


\author{C. Ian Short}
\affil{Department of Astronomy \& Physics and Institute for Computational Astrophysics, Saint Mary's University,
    Halifax, NS, Canada, B3H 3C3}
\email{ian.short@smu.ca}




\begin{abstract}

  We present an alternate deployment of the GrayStarServer (now ChromaStarServer (CSS)) 
pedagogical stellar atmosphere and spectrum synthesis WWW-application,
namely ChromaStarDB (CSDB), in which the atomic line list used for spectrum synthesis is implemented
as an SQL database table rather than as a more conventional byte-data file.  This allows for
very flexible selection criteria to determine which transitions are extracted from
the line list for inclusion in the synthesis, and enables novel pedagogical and research experiments in
spectrum synthesis.  This line selection flexibility is reflected in the CSDB UI.  The database
extraction is very fast and would be appropriate for the larger line lists of research-grade
modeling codes.  We also take the opportunity to present major additions to the ChromaStar
and CSS codes that are also reflected in CSDB: i) TiO band opacity in the JOLA approximation,
ii) Metal $b-f$ and Rayleigh scattering opacity, iii) 2D implementation of the flux integral,
iv) Improvement of the $n_{\rm e}$ convergence, v) Expansion of the exo-planet modeling
parameters, vi) A red giant template model for the initial guess at the structure, and 
vii) General improvements to the UI.  
The applications may be found at the home page of the OpenStar project: www.ap.smu.ca/OpenStars/. 
   
\end{abstract}



\section{Introduction}

 In \citet{gss16} (S16 henceforth) and papers in that series we describe GrayStar and GrayStarServer, two novel applications
that implement approximate physical stellar atmosphere and spectrum modeling in JavaScript (JS) and Java, respectively.  The applications have
a pedagogically intuitive and engaging 
user interface (UI) and visualization component written in HTML and JS that students and instructors can access through any WWW 
browser on any commonplace
personal computing device as equipped by default.  This virtual apparatus allows instructors to engage in demonstration-based 
teaching in class, as found effective by physics education research (PER), and to assign lab projects at a wide range of pedagogical levels,
as well as allowing for modeling in support of research that is carefully limited in scope.   
 The broader significance is that now that the feasibility of responsive modeling that is realistic enough for a range of
pedagogical applications has been demonstrated for commonplace platforms equipped only with a WWW browser, it only remains
for experts in UI design and web programing to provide an even more intuitive and compelling UI.  These applications suggest 
the possibility that as commonplace devices and the web become more capable, there may be an increasing integration of
research and EPO modeling that might have significant consequences for the public appreciation of astronomy and astrophysics and
the recruitment into university astronomy programs, particularly if researchers in other areas of computational astrophysics
also deploy web-based didacticized versions of their modeling and visualization.

\paragraph{}
 
 The most important distinction between GrayStar and GrayStarServer is
that the former carries out all the modeling in JS on the client-side device and only includes spectral line
opacity for about 20 spectral lines that are most important for spectral classification at MK spectral dispersion, 
whereas the latter carries 
out the physical modeling in Java on the server-side, and can thus perform and display 
spectrum synthesis based on a more comprehensive atomic line list of thousands of lines, and 
thereby expands the range of projects
that can be carried out (see S16).  S16 also included a comparison of synthetic spectra computed
with GrayStarServer and version 15 of Phoenix \citep{phoenix} for select early- and late-type atomic spectral
diagnostics for select stellar parameters.

\paragraph{}

  Here we introduce ChromaStarDB (CSDB), an alternate deployment of GrayStarServer that implements the atomic line list
 as a database that is accessible with structured query language 
(SQL), thereby making it straightforward to implement code that provides for pedagogical spectrum synthesis experiments with subsets of atomic transitions that are
selected according to flexible atomic and transition data criteria.  In Section \ref{srev} we review the overall modeling strategy and major approximations being
adopted, and the line list database development is described in Section \ref{ssql}.  We also report three significant developments that have 
been made to GrayStar and GrayStarServer, and that are also reflected in CSDB,
the most important of which are: 1) Molecular band opacity arising from TiO,
 2) Opacity from the photo-ionization of ``metal'' species and from Rayleigh scattering,
and 3) 2D flux integral evaluation and proper macroscopic broadening, 
which are described in Sections \ref{sTiO}, \ref{sopac}, and \ref{s2dflux}, respectively.  These changes make the physical 
modeling more realistic and general, and expand the range of topics that can 
be credibly demonstrated and investigated.  Additionally there are several changes that have an impact 
on the usefulness and range of the applications, including expansion of the exo-planetary modeling parameters
to include atmospheric surface pressure and choice of organic solvent, and greater control over the 
compromise between modeling realism and execution time, as 
described in Section \ref{sminor}.  In Section \ref{sdiscuss} we conclude with a brief discussion of the 
significance and context.  As a point of nomenclature, we note here that GrayStar and GrayStarServer have been re-named 
ChromaStar (CS, Astrophysics Source Code Library ascl:1701.008) and ChromaStarServer (CSS, ascl:1701.009) 
since our last published report to
reflect the more realistic {\it non}-gray treatment of the opacity distribution and the $T_{\rm kin}(\tau)$ structure that those codes have now had since 2016, as
described in S16, and will be referred to as such henceforth in this work.  All applications may be found 
at the home page of the OpenStar project (www.ap.smu.ca/OpenStars).

\section{Review of strategy and approximation \label{srev}}

  The fundamental approximation upon which CS and CSS are based is that the $T_{\rm kin}(\tau)$ structure is 
obtained by re-scaling that of a research-grade model, $T_{\rm kin, 0}(\tau)$, computed with V. 15 of Phoenix of either an 
early- or late-type template
star (see Section \ref{stemplate} for more detail and a discussion of how this has been improved) as
$T_{\rm kin}(\tau) = (T_{\rm eff}/T_{\rm eff, 0}) T_{\rm kin, 0}(\tau)$.  This obviates the need for
treating the thermal radiative equilibrium problem, and thus
of treating spectral line blanketing opacity (it is for this reason that the microturbulent velocity
dispersion is not an atmospheric structure modeling parameter in this approach). 
 As a result, we only need the value of the specific intensity,
$I_\lambda(\tau, \cos\theta)$, at the surface, $\tau = 0$, and only at those $\lambda$ points required for
the presentation of the overall spectrum or of spectral features.  This one approximation allows us
to forgo the most intensive part of the atmospheric modeling problem, and is a surprisingly good
approximation.  The gray solution for $T_{\rm kin}(\tau_{\rm gray})$ on a gray optical depth scale, $\tau_{\rm gray}$,
shows that we should expect the $T_{\rm kin}(\tau)$ values to vary approximately linearly with $T_{\rm eff}$ (see \citep{gray}).  
The resulting model gives rise to emergent observables that are credible at at least the level of 
visual inspection.

\paragraph{}

Initial guesses for the gas pressure ($P_{\rm gas}(\tau)$) and electron pressure ($P_{\rm e}(\tau)$) structures are calculated by 
re-scaling the corresponding structures from one of the template models with, variously,  $T_{\rm eff}$, 
$\log g$, metallicity ($[M/H]$), 
and helium abundance ($[He/H]$) according to Ch. 9 of \citet{gray}.  The code then refines these structures by
iteration among the formal solution for the hydrostatic equilibrium (HSE) equation to obtain the 
total pressure ($P(\tau)$) structure, the 
equation of state (EOS) for a partially ionized ideal gas to obtain the mean molecular weight ($\mu(\tau)$)
 and mass density ($\rho(\tau)$) structures, and the extinction coefficient
distribution, $\kappa_\lambda(\tau)$.  The refined $P_{\rm gas}(\tau)$ values are recovered by subtraction 
of the radiation pressure, $P_{\rm rad}(\tau)$, from the total $P(\tau)$ values, under the assumption
that the $P_{\rm rad}$ distribution is that of a radiation field with an ideal blackbody spectrum.  This outer iteration incorporates an inner iteration
between the electron density ($n_{\rm e}(\tau)$) structure and the ionization fractions of all elements up 
to and including those of the Fe-group. 
 However, for both the
outer and inner iterations, we do not attempt to meet a convergence criterion, but only 
iterate for a specified number of times (see Section \ref{srealism}).   
Generally, the procedures used are those as described in Chs. 8 and 9 of \citet{gray}, and 
the application can be viewed as a pedagogical supplement to that reference.  
The overall procedure executes of the order of 10 iterations in under a few seconds on a 
commonplace laptop once the
JS code has been initially loaded and interpreted.

\paragraph{}

  The overall spectrum is sampled at only 350 wavelength points, currently distributed from
260 to 2600 nm in equal intervals of $\log\lambda$, which suffices to present the shape of the
spectral energy distribution (SED).  This background $\lambda$ grid is supplemented, {\it ad hoc},
by local grids centered on each spectral line that meets the 
$\kappa^{\rm l}_\lambda(\tau) / \kappa^{\rm c}_\lambda(\tau)$ 
criterion at any of three $\tau$ values for inclusion in the total $\kappa_\lambda$ spectrum, where 
$\kappa^{\rm l}_\lambda$ and $\kappa^{\rm c}_\lambda$
denote the line and background continuum opacities, respectively.  Each of these local $\lambda$ grids 
samples the Gaussian line core with nine $\lambda$ points spaced equally in ($\lambda - \lambda_0$),
 where $\lambda_0$ is the wavelength of line center, and with 20 $\lambda$ points (10 per wing) that sample the Lorentzian (or linear Stark) 
 wings 
in equal intervals of $\log(|\lambda - \lambda_0|)$ (see \citet{methods} for more detail), for a total
of 29 $\lambda$ points per line.

\section{SQL database-driven spectrum synthesis \label{ssql}}

  As described in S16, CSS reads as input a limited list of atomic line transitions selected from the NIST 
Atomic Spectra Database \citep{nist}.  The line list is accessed in the conventional way by performing
a file I/O byte-array read operation.  Although much faster than a character-read operation on an ASCII
file, this method still has the disadvantage that {\it every} record must be read from the line list
file and any selection criteria based on the atomic data values must be performed as a separate operation
upon all records after the file I/O operation.  This is not a concern for normal research-oriented
spectrum synthesis where normally the only line-wise selection criterion is based on a trial line opacity
calculation and requires the transition data in any case.  However, for pedagogical spectrum synthesis
it would be insightful to select which transitions will be included according to a broader set of criteria. 
With conventional file I/O, in addition to the time wasted reading unnecessary records, this would require 
cumbersome multiply nested 
\texttt{if-then-else} logic to apply multiple selection criteria based on atomic data to line transitions.

\paragraph{}

  By contrast, the CSDB line list has been stored as a MySql database table and the Java spectrum
synthesis procedure accesses it with an SQL \texttt{SELECT} query that only extracts those
records that satisfy the parameters of the query.  Those parameters are
determined by values and selections made by the user with the CSDB UI.  
 On-line spectral line
data archives have already been implemented as SQL-type databases, and
CSDB incorporates the approach much more directly in a spectrum synthesis code.
The database extraction is at least
as fast as buffered byte-data array file I/O, and would be feasible for the larger atomic and molecular line
lists used by research-grade modeling codes. 
 The interaction between
Java and the local MySql server is enabled by the \texttt{Connector/J} Java Database 
Connectivity (JDBC) driver
provided by MySql.  This capability
exemplifies one main advantage of implementing an astrophysical modeling code in a contemporary
programming language - that well established utilities are readily available for interacting with
other modern computing tools, such as SQL databases.  

\paragraph{}

Currently, the
selection criteria controllable by the user include

 \begin{enumerate}

  \item{Chemical element (H ($Z=1$) through Zn ($Z=30$) and Ge ($Z=32$), and the relatively
detectable $n-$capture elements Rb ($Z=37$), Sr ($Z=38$), Cs ($Z=55$) and Ba ($Z=56$)).} 
  \item{Ionization stage (I through IV).}
  \item{Minimum and maximum oscillator strength, $\log f$ (-5.0 to 2.0).}
  \item{Minimum and maximum excitation energy of the lower energy level of the transition, $E_{\rm i}$ (0.0 to 45.0 eV).}
  \item{Wavelength range, $[\lambda_1$, $\lambda_2]$, within the specified synthesis range.}       

 \end{enumerate}

\begin{figure}
\plotone{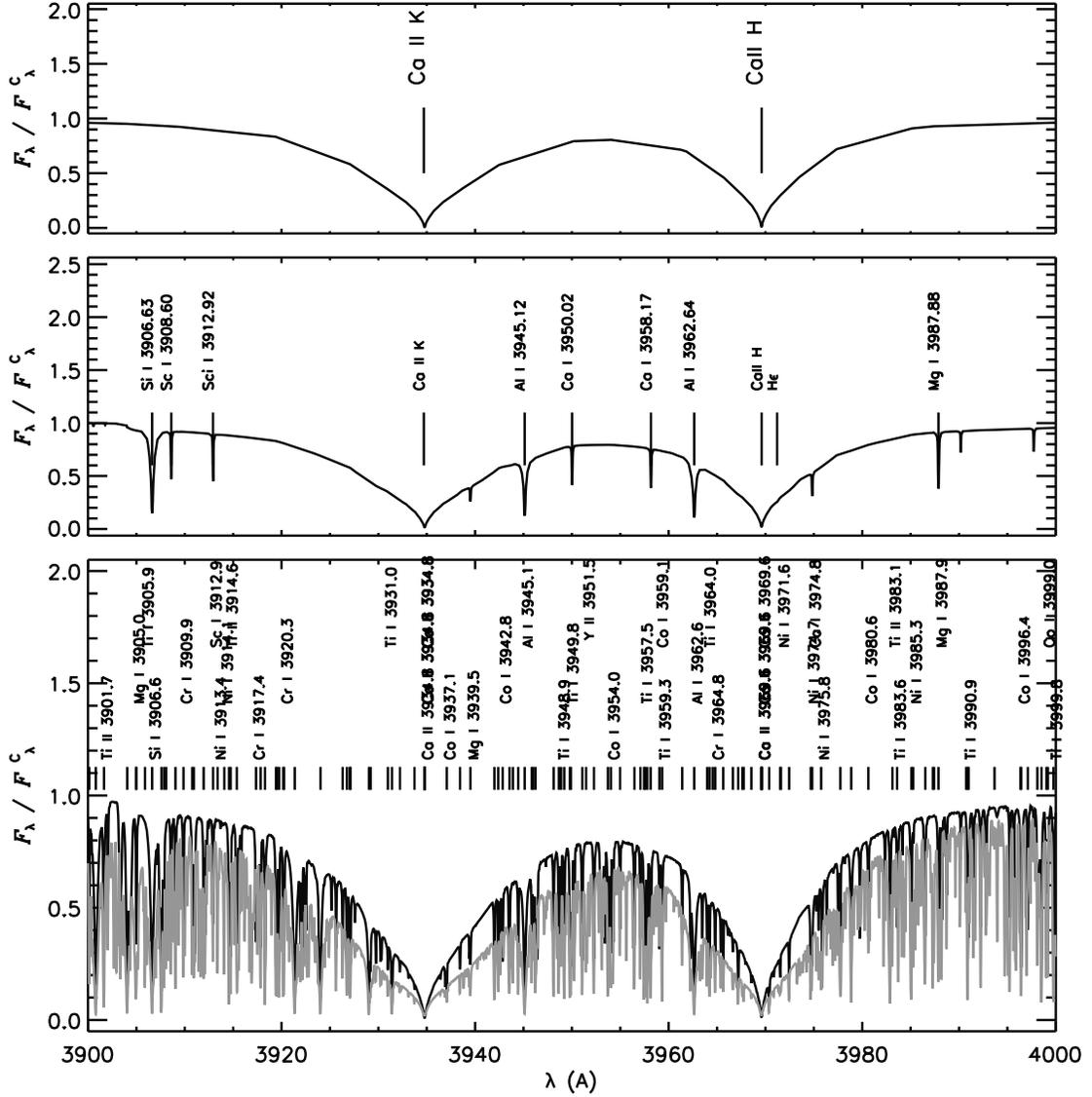}
\caption{Series of synthetic spectra computed with CSDB for the \ion{Ca}{2} HK region 
for the Sun with various selection criteria for the MySql database line extraction:
i) \ion{Ca}{2} only (upper panel); ii) Criterion i), \ion{H}{1} and light metals 
(middle panel), and iii) Criteria i) and ii) and Fe-group elements (lower panel).  
For comparison, we have overplotted the Phoenix synthetic spectrum in the lower
panel (gray line) - see text.  Line identifications are those of CSDB in the upper and
middle panel, and those of Phoenix in the lower panel.  To manage the over-crowding of 
spectral line labels 
in the lower panel we have omitted the labels for all \ion{Fe}{1} and II lines.     
  \label{fCaIIHK}
}
\end{figure}

  These selection criteria allow, for example, pedagogical spectrum synthesis experiments in which a 
crowded diagnostically important spectral region is synthesized in stages with successive
groups of species being included in the synthesis, as illustrated for the important
\ion{Ca}{2} HK region in the Sun in Fig. \ref{fCaIIHK},
where we also overplot the synthetic spectrum as computed with version 15 of Phoenix for comparison
(note that we do not necessarily expect a close match to Phoenix results because
of the approximate nature of CSDB and the sensitivity of saturated line wings 
to the background opacity).  
For example, this region could be synthesized first with only \ion{Ca}{2}
selected, then with \ion{H}{1} and light metals included to illustrate the blending 
of \ion{Ca}{2} H with H$\epsilon$ in an uncrowded synthesis among other things, and 
then with the much larger
number of lines contributed by the Fe-group elements included.  
Inversely, one could 
gain insight into a crowded region in a late-type star by excluding the numerous \ion{Fe}{1} lines only. 

\section{TiO band opacity \label{sTiO}}

\begin{figure}
\plotone{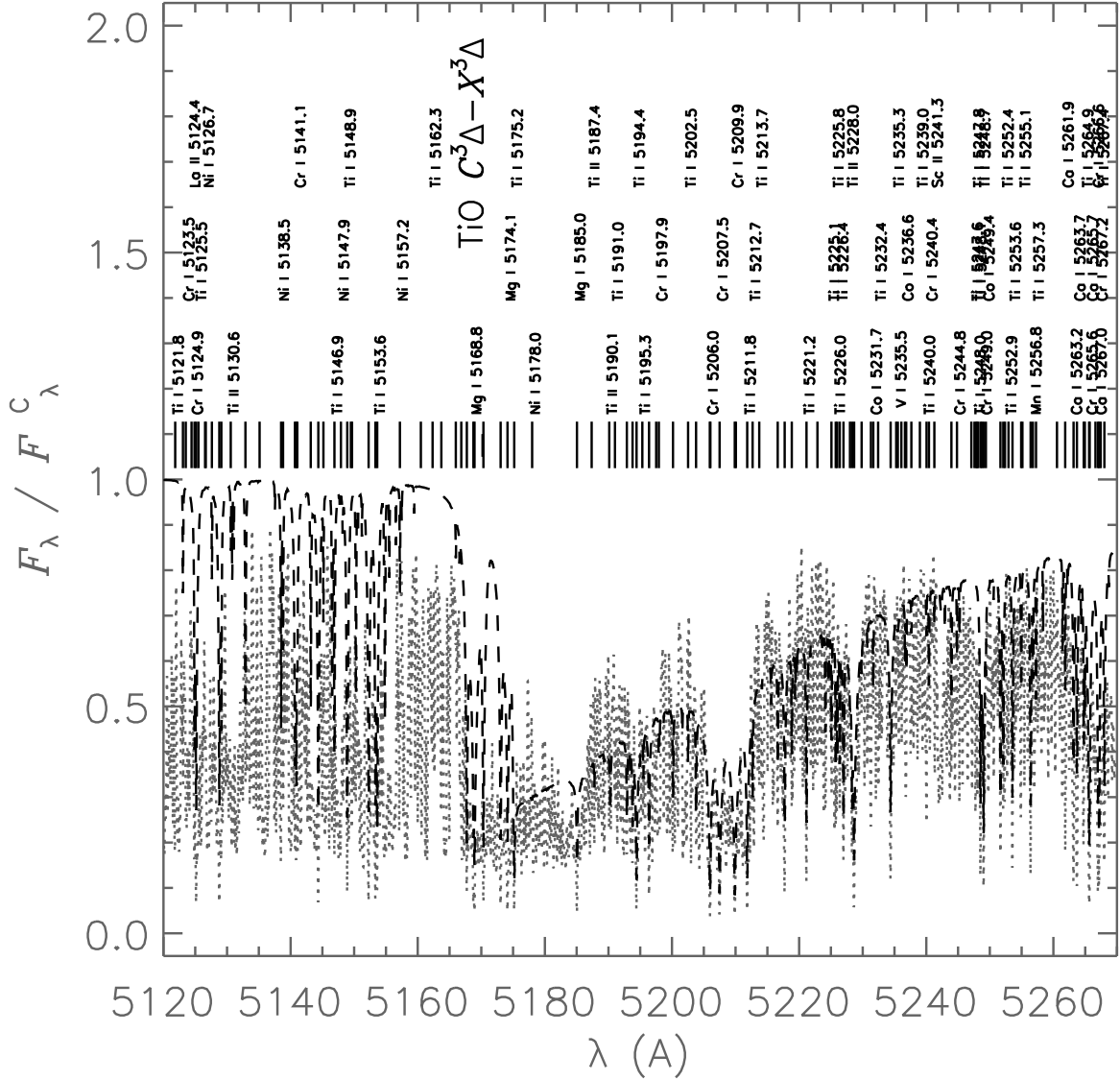}
\caption{
 TiO band origin region for the $C^3\Delta - X^3\Delta$ transition ($\lambda_{00}=517.02$ nm) 
for a solar metallicity model of $T_{\rm eff} = 3750$ K and $\log g = 4.5$ as computed with 
the JOLA approximation by CSS (dashed line) and in detail with Phoenix (dotted line).  
Line identifications are those of Phoenix spectrum synthesis and to reduce overcrowding
of line labels, we have omitted labels for all \ion{Fe}{1} and II lines.    
  \label{fTiO3750455170}
}
\end{figure}

\begin{figure}
\plotone{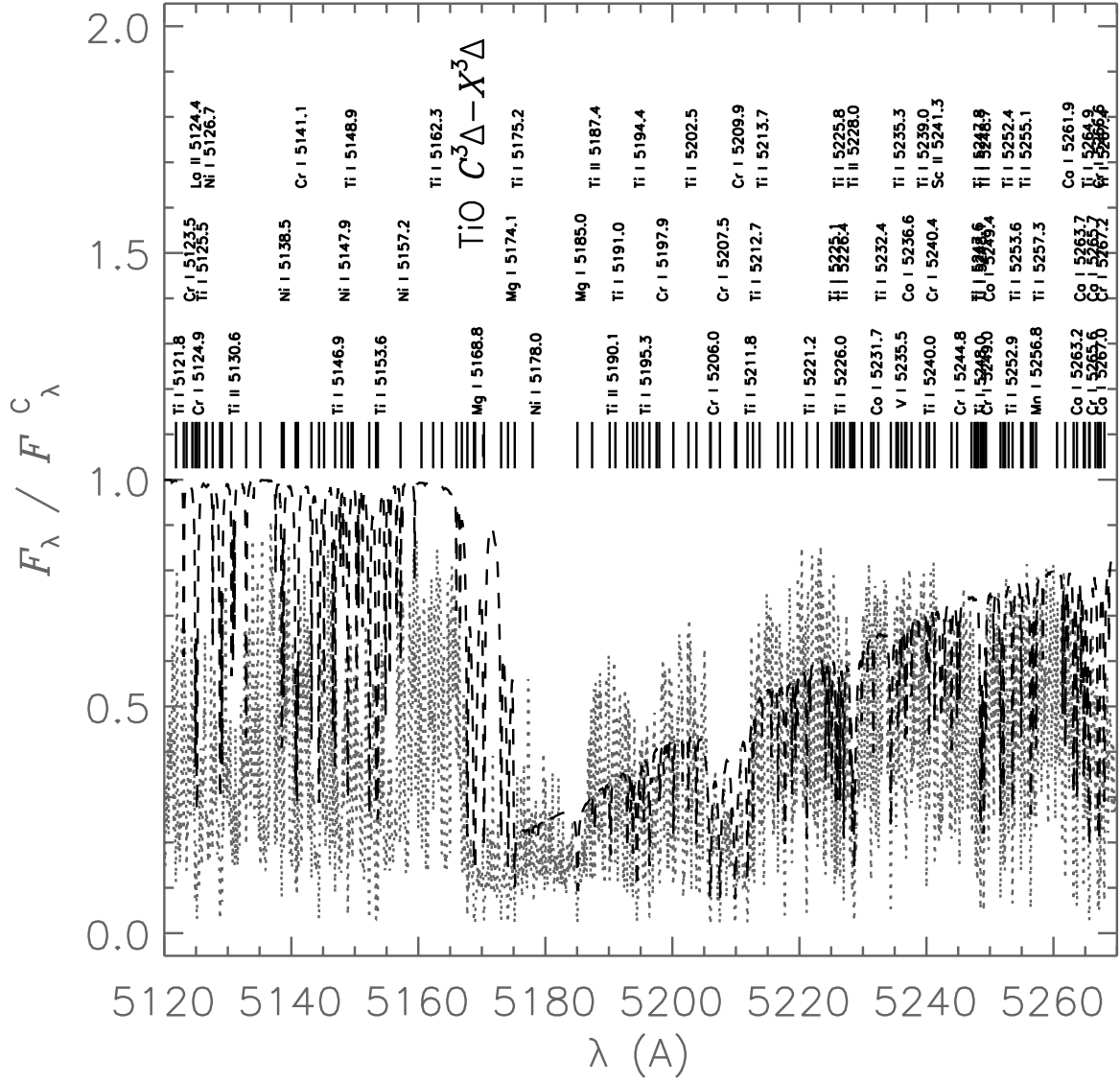}
\caption{Same as Fig. \ref{fTiO3750455170} except for a model of $\log g = 2.0$.
  \label{fTiO3750205170}
}
\end{figure}

\begin{figure}
\plotone{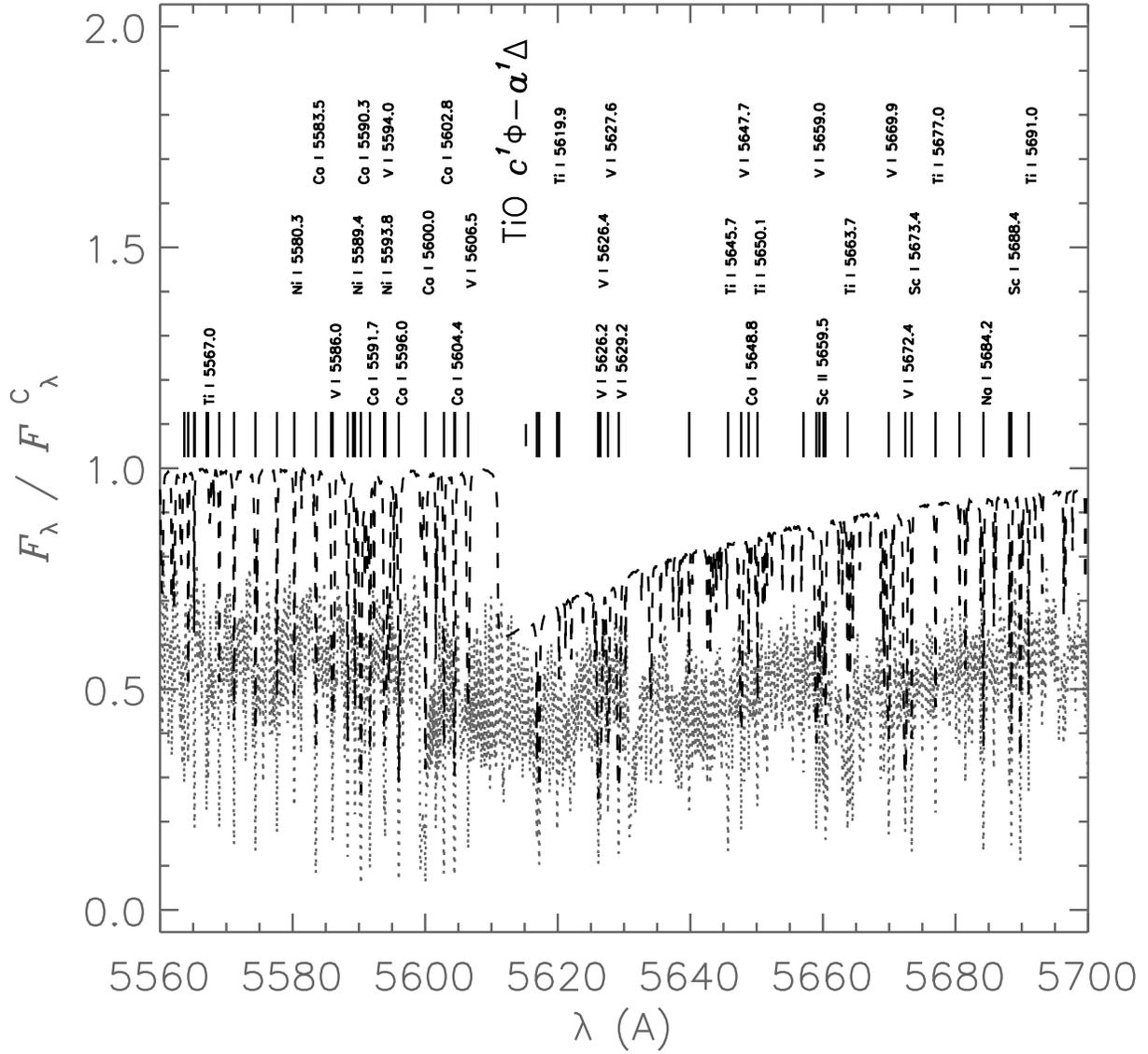}
\caption{Same as Fig. \ref{fTiO3750205170} except for the $c^1\Phi - a^1\Delta$ transition ($\lambda_{00}=560.52$).
  \label{fTiO3750205605}
}
\end{figure}

\begin{figure}
\plotone{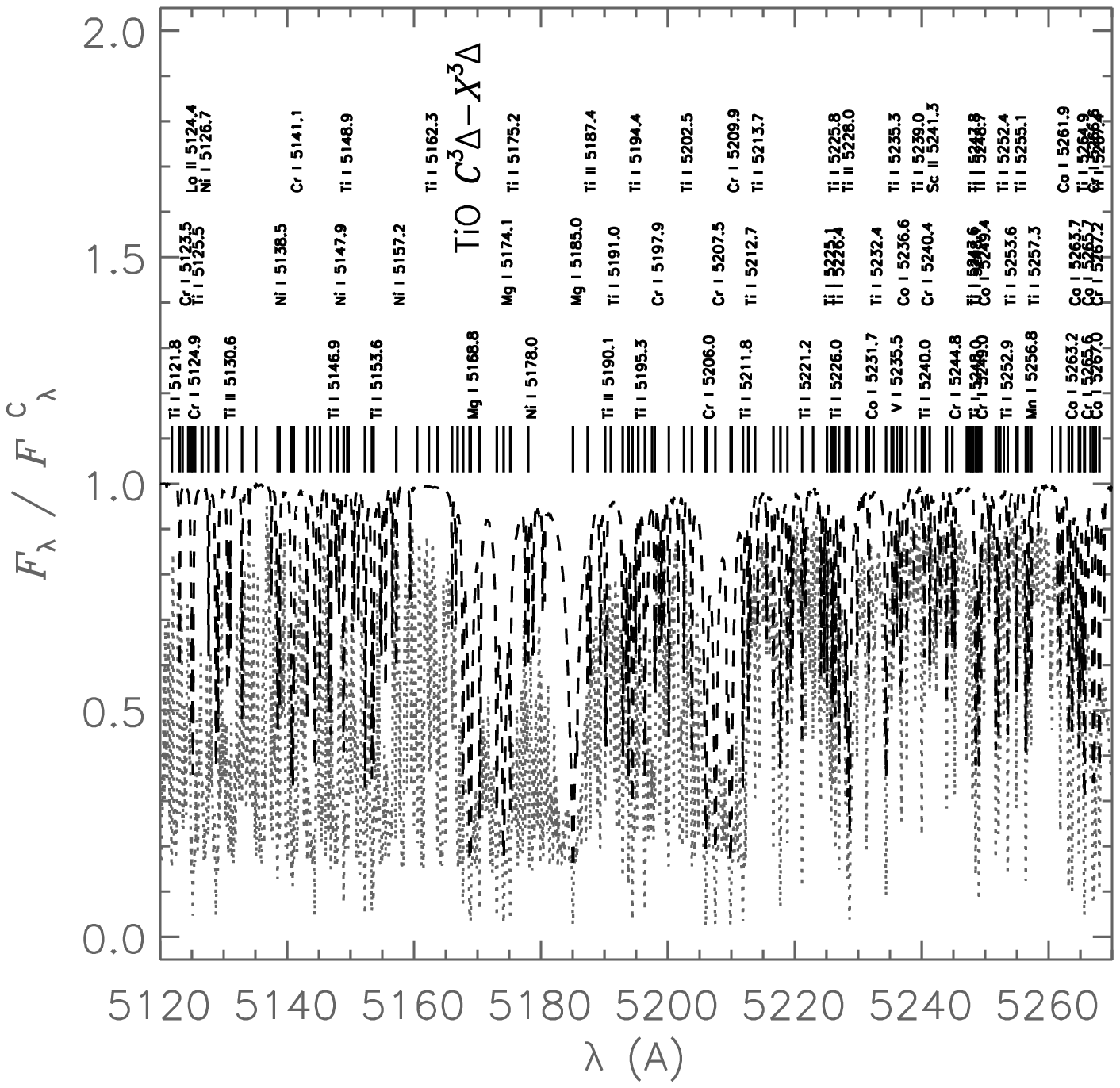}
\caption{Same as Fig. \ref{fTiO3750205170} except for a model of $T_{\rm eff} = 4250$ K and $\log g = 2.0$.
  \label{fTiO4250205170}
}
\end{figure}

  We have added to the opacity calculation in CS, CSS, and CSDB 
for stars of $T_{\rm eff} < 5000$ K
the contribution from the vibronic transitions of the {\it P} and {\it R} 
branches of
three electronic bands of TiO falling in the visible or near IR spectral regions: 
$C^3\Delta - X^3\Delta$ ($\alpha$ system, band origin, $\lambda_{00}=517.02$ nm),
 $c^1\Phi - a^1\Delta$ ($\beta$ system, $\lambda_{00}=560.52$ nm), 
and $A^3\Phi - X^3\Delta$ ($\gamma$ system, $\lambda_{00}=709.43$ nm).  The first two
of these are important MK spectral classification diagnostics for class {\it M} stars. 
The large number of individual transitions associated with molecular bands is 
prohibitive to treat in detail for the low performance platforms for which CS, 
CSS, and CSDB were designed.  Therefore,
the TiO opacity is treated with the Just-Overlapping-Line-Approximation (JOLA) as formulated
by \citet{jola}, 
 and sampled with 100 equally space $\lambda$ points. 
Values for most molecular data were taken from \citet{allens}, and  
we set the scale of the band strengths by matching the
strength of the TiO bands as treated in detail ({\it ie.} line-by-line) by Phoenix for a model 
of $T_{\rm eff} = 3725$ K and $\log g = 2.0$ 
 (see Figs. \ref{fTiO3750455170} to \ref{fTiO4250205170}).  

\paragraph{}

  The TiO JOLA band opacity procedure requires as input the number density of TiO molecules
at each depth in the atmospheric model, $n_{\rm TiO}(\tau)$.  For stars of $T_{\rm eff} < 7300$ K,
 we iteratively solve a set of 
explicitly coupled equilibrium equations for the number density fractions $n_{\rm TiO}/n_{\rm Ti}$, 
$n_{\rm Ti I}/n_{\rm Ti}$, and $n_{\rm O I}/n_{\rm O}$, where the denominators are stipulated
by the input abundances, along with similar fractions for other important $e^-$-donating elements.

These ratios are evaluated in terms of the quantities on the left-hand-side (LHS) of the 
Saha equation in close-to-standard form for diatomic molecules and ions, or the reciprocals thereof, 
with either the number density of one of the reactant atoms ($n_{\rm A}$) or $n_{\rm e}$, respectively, moved
to the right-hand-side (RHS) and set equal to its previous value in the iteration (or its initial value).  
As an illustrative example for the simplified case of only three ionization stages of Ti and two of O,
this gives:

\begin{equation}
n_{\rm O I}/n_{\rm O} = 1.0 / ( 1.0 + n_{\rm O II}/n_{\rm O I} + n_{\rm TiO}/n_{\rm O I} ),
\label{eqSahaO}
\end{equation}

\begin{equation}
n_{\rm Ti I}/n_{\rm Ti} = 1.0 / ( 1.0 + n_{\rm Ti II}/n_{\rm Ti I} + (n_{\rm Ti III}/n_{\rm Ti II})(n_{\rm Ti II}/n_{\rm Ti I}) + n_{\rm TiO}/n_{\rm Ti I} ),
\label{eqSahaTi}
\end{equation}

\begin{equation}
n_{\rm TiO}/n_{\rm Ti} = (n_{\rm TiO}/n_{\rm Ti I}) / ( 1.0 + n_{\rm Ti II}/n_{\rm Ti I} + (n_{\rm Ti III}/n_{\rm Ti II})(n_{\rm Ti II}/n_{\rm Ti I}) + n_{\rm TiO}/n_{\rm Ti I} ) 
\label{eqSahaTiO}
\end{equation}

In Eq. \ref{eqSahaTiO} we could have chosen to evaluate the ratio $n_{\rm TiO}/n_{\rm O}$ instead, but choose the atomic reactant that couples to the fewest
other molecules as the denominator to simplify the coupling among molecules that share reactants when we expand the network to
other molecules.  The value of $n_{\rm O I}$ or $n_{\rm e}$ being used in the Saha equation for either 
$n_{\rm TiO}/n_{\rm Ti}$
or ionic ratios such as $n_{\rm Ti II}/n_{\rm Ti I}$, respectively, is always that of the previous iteration.
By default this set is iterated eight times to achieve an approximately self-consistent solution
to the coupled molecular- and ionization equilibria and $n_{\rm e}$ values.  The default value of
eight iterations is an approximate minimum required to damp out significant changes in the 
number densities from one iteration to the next for most late-type stars, as determined by
numerical experimentation.  However, the user can vary the number of iterations from five to 12 
in the new ``Performance/Realism'' panel of the CS, CSS, and CSDB UIs, and thereby manage the compromise between 
execution time and modeling realism as appropriate for the particular context (see also Section \ref{sNe}). 
The user can choose to display the $n_{\rm TiO}(\tau)$ distribution in the optional advanced
output (see S16).  Currently, CS, CSS, and CSDB increasingly over-estimate the value of $n_{\rm TiO}(\tau)$, and, hence,
the strength of the TiO bands, as $T_{\rm eff}$ decreases because other oxide molecules that compete
for \ion{O}{1} particles, particularly CO, are not yet included in the ionization-chemical equilibrium
treatment.  

\paragraph{Pedagogical application }

Although
bands treated with JOLA appear crude in the plot of the synthetic spectrum,
they do look realistic when viewed in the direct rendering of the spectral image
at MK classification dispersion featured in the CS, CSS, and CSDB UIs.   
Therefore, their addition makes the applications much more appropriate for pedagogical 
demonstrations and investigations of late {\it K} and early {\it M} class stars.  The user has the option of
deselecting the TiO bands to remove them from both presentations of the spectrum
as an aid to clarity when studying the behavior of atomic lines. 
An example of a pedagogical activity enabled by the addition of the JOLA bands is
to have students estimate both the value, and the uncertainty of, $T_{\rm eff}$
for which the TiO band heads just become visible in the rendering of the spectral image (see S16),
 and to explore the dependence of that $T_{\rm eff}$ value on $\log g$.
This would give students
at least an approximate sense of how molecular equilibrium depends on both $T_{\rm kin}$ and
$P_{\rm gas}$, and of why the $T_{\rm eff}$-MK-spectral-class relation is $\log g$-dependent.

\section{Metal $b-f$ and Rayleigh scattering opacity \label{sopac}}

 In addition to the direct effect on the computed spectrum, the continuous opacity calculation also affects the
atmospheric structure because we are integrating the formal solution of the HSE equation on the $\lambda$-averaged 
Rosseland optical depth scale, $\tau_{\rm Ros}$ (see Section \ref{srev}), and the corresponding mean extinction coefficient, $\kappa_{\rm Ros}(\tau)$, 
explicitly appears on the RHS of the expression for the formal solution. 

  As of S16, CS and CSS included continuous extinction arising from photo-ionization (bound-free, $b-f$) and
bremstrahlung (free-free, $f-f$) transitions, variously, of \ion{H}{1}, H$^-$, H$_{\rm 2}^+$, \ion{He}{1}, \ion{He}{2}, 
and He$^-$ and Thomson scattering by free $e^-$ particles.  With only these source, the total continuous extinction
is expected to be increasingly underestimated with decreasing wavelength below about 400 nm where both $b-f$ extinction
by metals and Rayleigh scattering become increasingly important.  As a result, the spectrum synthesis capability of CSS 
was restricted to $\lambda > 360$ nm.  

\paragraph{}

  We have incorporated $b-f$ extinction from one or more atomic $E$-levels, including the ground state, arising from seven 
``metal'' species that are known to be especially 
important opacity sources, namely \ion{C}{1}, \ion{Mg}{1}, \ion{Mg}{2}, \ion{Al}{1}, \ion{Si}{1}, \ion{Si}{2}, and \ion{Fe}{1}.  
The procedures are adapted from those of the ``JUL2014'' release of the Moog spectrum synthesis code \citep{moog}, ported from Fortran to 
Java (CSS and CSDB) and JS (CS), and adapted by incorporating the correction for stimulated $b-f$ emission 
(stimulated photo-recombination)
in local thermodynamic equilibrium (LTE).  The latter is a small correction, but because the CS and CSS source code files
are pedagogical documents, we chose to include it.  The Moog implementation of these opacity routines is in turn
based on the opacity procedures of the Atlas9 atmospheric modeling code \citep{atlas9}, and the various laboratory data
on which they are based are described in papers in that series.  These procedures require as input the ground state
population of the relevant species at each depth in the atmospheric model, and that is a quantity that CS and CSS (and
now CSDB) compute as of S16 to compute line (bound-bound, $b-b$) opacity for the spectrum synthesis.

\begin{figure}
\plotone{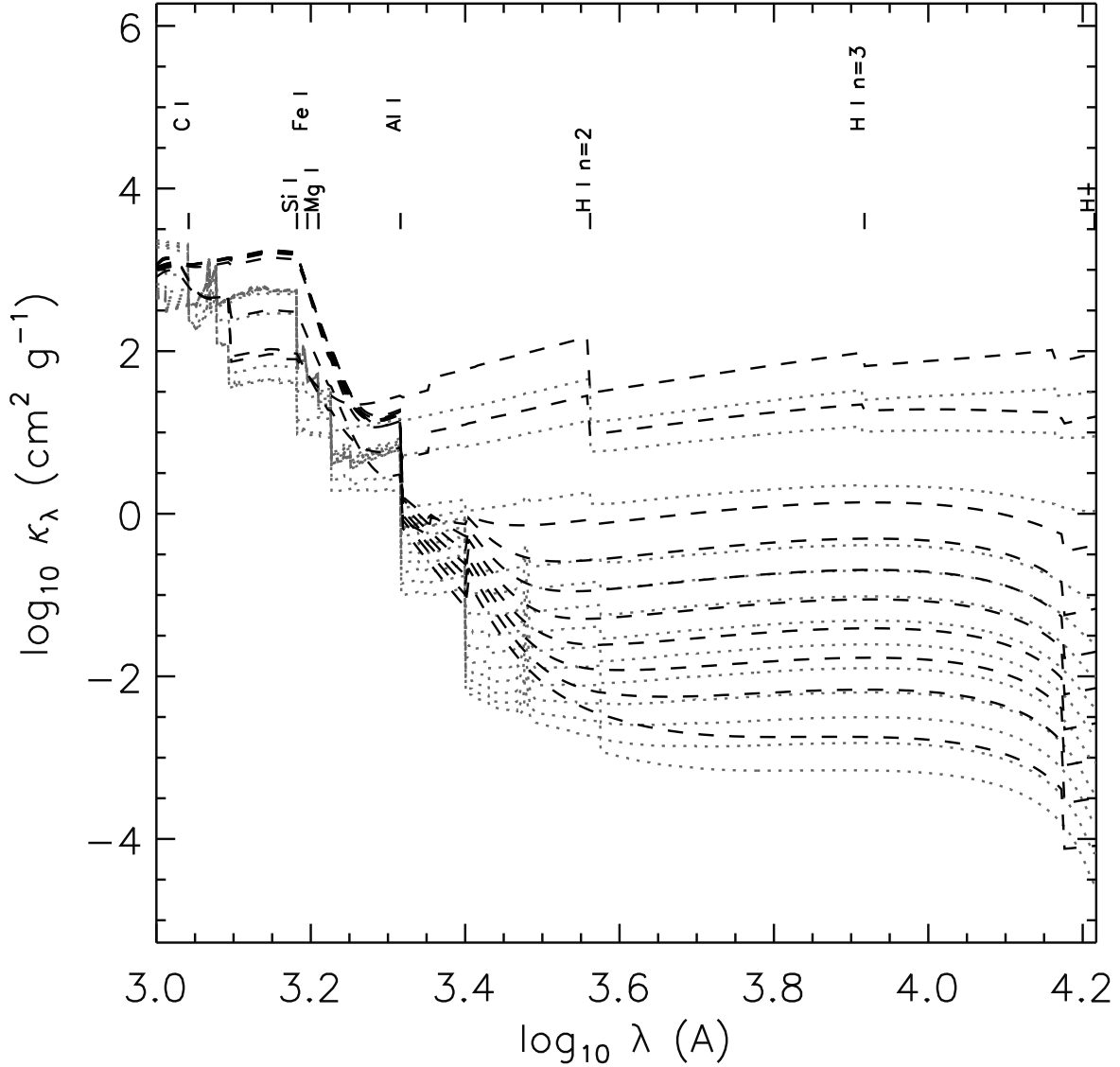}
\caption{
Mass extinction coefficient distribution, $\kappa_\lambda(\lambda)$ for a selection of
$\log\tau$ values spanning the total thickness of the atmosphere in approximately equal
decades 
for a solar metallicity model of $T_{\rm eff} = 5000$ K and $\log g = 4.5$ as computed with 
CSS (dashed line) and with Phoenix (dotted line).  The $\log\lambda_{\rm o}$ values of 
ground state $b-f$ transitions of metals and other $b-f$ transitions of H are indicated.  
No attempt has been made to match $\log\tau$, $T_{\rm kin}$, or $P_{\rm gs}$ values
at which $\kappa_\lambda(\lambda)$ has been plotted from each code - see text.   
  \label{fOp500045}
}
\end{figure}

\paragraph{}

Similarly, we have adapted and incorporated routines from Moog for the Rayleigh scattering opacity arising from
\ion{H}{1} and \ion{He}{1}.  Moog also provides a procedure for Rayleigh scattering by H$_{\rm 2}$, but we must defer
implementation until CS, CSS, and CSDB are able to compute the population of H$_{\rm 2}$.  Currently, TiO is 
the sole molecule being treated, but we plan an aggressive expansion of the molecular network now that our 
implementation of the unified molecular and ionization equilibrium has been proven.  Rayleigh scattering from
these species is a minor, but non-negligible, contributor to the continuous extinction in late-type stars. 
Fig. \ref{fOp500045} shows a comparison of the mass extinction distribution, $\kappa_\lambda(\lambda)$, for
a selection of $\log\tau$ values spanning the thickness of the atmosphere in approximately equal 
$\log\tau$ intervals as computed by CSS and by Phoenix.  No attempt has been made to match the $\log\tau$,
$T_{\rm kin}$ or $P_{\rm gas}$ values at which the $\kappa_\lambda(\lambda)$ distributions from each 
code are being displayed.  However, the plot gives an overview of how the variation of $\kappa_\lambda(\lambda)$
as a function of $\lambda$ and $\tau$ as computed by CSS compares to that of Phoenix.  We note that we do not expect the
$\kappa_\lambda(\lambda)$ distributions of the two codes to have the same qualitative features because 
version 15 of Phoenix adopts the resonance averaged Opacity Project ground state $b-f$ cross-sections of \citet{bautista}, as 
described in \citet{shortPHX}.

\section{2D flux integral \label{s2dflux}}

  As of S16, CS and CSS calculated the surface flux, $F_\lambda(\tau=0)$, by integrating the specific surface intensity, 
$I_\lambda(\tau=0, \cos\theta, \phi)$,
over annular $\Delta\cos\theta$ elements centered on the substellar point ($\cos\theta=1$), following standard procedure for 
$I_\lambda(\cos\theta, \phi)$ fields that are expected to be axi-symmetric about the positive $z$ axis,
which is normally defined to project from the substellar point to the observer and defines the polar direction $\theta = 0$
({\it ie.}, the assumption that $I_\lambda(\cos\theta, \phi) = I_\lambda(\cos\theta)$.)  This treatment simplifies the 
$F_\lambda(\tau)$ calculation by reducing a two-dimensional numerical integration over ($\cos\theta$, $\phi$) to a one-dimensional
integration over $\cos\theta$.  It also implies horizontal
homogeneity of the atmosphere ($Q(\tau, x, y) = Q(\tau)$ for all atmospheric quantities, Q, where the radial optical depth, $\tau$,
 axis is anti-parallel 
to $z$).  More practically, this
approach forces the treatment of macroscopic broadening effects on the $F_\lambda(\tau=0)$ spectrum to be applied
{\it post hoc} by convolution of $F_\lambda(\tau=0)$ with the appropriate broadening kernel functions, as 
implemented in CS and CSS as of S16.  This approach is especially awkward for CS, CSS, and CSDB because the 
 $\lambda$ sampling of the spectrum is highly irregular to expedite the spectrum synthesis (see Section \ref{srev}). 
 As a result, an
awkward interpolation of this irregularly and sparsely sampled $F_\lambda(\tau=0, \lambda)$ distribution onto a
fine uniformly sampled $\lambda$ grid was necessary to prepare the spectrum for {\it post hoc} convolution
(see S16).  

\paragraph{}

  CS, CSS, and CSDB now compute the $F_\lambda(\tau=0)$ distribution by performing the double numerical integration
over ($\Delta\cos\theta$, $\sin\theta\Delta\phi$) solid angle elements.  In addition to the 11 $\cos\theta$ values distributed with a Gauss-Legendre 
quadrature, each annulus is now divided into 36 uniformly sampled $\Delta\phi$ intervals, yielding nine $\Delta\phi$ elements per 
quadrant of the unit circle centered on the substellar point.  The direction of $\phi = 0$ is taken
to be the direction of the positive $x$ axis such that, along with the $z$ axis, a right-handed 3D Cartesian coordinate 
system with an origin, $O$, at the substellar point is defined and $\phi$ increases counter-clockwise.  Thus the projected 
visible stellar hemisphere is sampled with 396 ($\Delta\cos\theta$, $\sin\theta\Delta\phi$) elements.  

\paragraph{}

The radial distance in the plane of the sky, $r(\cos\theta, \phi)$, from $O$ to each element is computed as $r = R\sin\theta$,
where $R$ is the radius of the star,
then projected onto the $x$ axis to arrive at a value of $x(\cos\theta, \phi) = r \cos\phi$.  The line-of-sight
component of the rotational surface velocity, $v_{\rm z, rot}(\cos\theta, \phi)$, is computed as $v_{\rm z, rot} = x v_{\rm eq}\sin i / R$,
where $v_{\rm eq}$ and $i$ are the surface equatorial rotation velocity and the inclination of the rotation axis with 
respect to the positive $z$ axis, respectively.
For each $(\Delta\cos\theta, \sin\theta\Delta\phi)$ element we generate a value of the line-of-sight component of the macro-turbulent velocity 
field, $v_{\rm macro}$, by randomly generating a number from a Gaussian distribution with a $\sigma$ value equal to the input macro-turbulent 
velocity parameter $\xi_{\rm macro}$.  The Gaussian distribution of random variables is produced by applying the 
polar form of the Box-Muller transformation \citep{boxmuller} to the uniformly distributed random variable in the interval $[0, 1]$ 
generated by the intrinsic Math.random() function in JS (CS) or Java (CSS and CSDB).  
For each $(\Delta\cos\theta, \sin\theta\Delta\phi)$ element we compute a total line-of-sight velocity, $v_{\rm z}$, arising from macroscopic broadening
processes by summing $v_{\rm z, rot}$ and $v_{\rm macro}$.  Each $I_\lambda(\cos\theta, \phi)$ beam is Doppler shifted 
according to $v_{\rm z}(\cos\theta, \phi)$, and the Doppler shifted $I_\lambda(\cos\theta, \phi)$ spectrum is interpolated onto
the observer's frame $\lambda$ grid and the shifted interpolated $I_\lambda(\cos\theta, \phi)$ spectra are summed to form
the broadened $F_\lambda$ spectrum.  

\paragraph{}

As before, the user controls the values of $v_{\rm z}$, $i$, and $\xi_{\rm macro}$ through 
the UI.  However, because only the Gaussian core of the lines are being sampled with equal $\lambda$ intervals
(see Section \ref{srev}), 
the maximum value of $v_{\rm eq}$ that can currently be accommodated is $\sim 20$ km s$^{-1}$ (values of $\xi_{\rm macro}$
are never greater than $\sim 8$ km s$^{-1}$ in any case).    
 CS does not perform spectrum synthesis with a more realistic line list as CSS and CSDB do, but it 
does provide a user-defined two-level atom (TLA) and computes and displays the one spectral line thereof (see \citet{methods}). 
An advantage of the {\it a priori} broadening based on the 2D flux integral is that this TLA spectral line in CS
can now also be straightforwardly subject to macroscopic broadening and we now provide that ability in the CS UI.
Additional potential advantages of the {\it a priori} broadening treatment are that it allows for i) The assumption
of horizontal homogeneity in the $F_\lambda$ spectrum synthesis to be relaxed and opens the way for an expansion of the UI
that allows users to add spots to stars and observe their effect on synthetic line profiles as treated in the
``1.5D'' approximation, and
ii) The adoption of radial-tangential macro-turbulence and, thus, the corresponding spectral line
bisector analysis \citep{grayfgk}. 

\section{Other additions and improvements \label{sminor}}

We take the opportunity here to report on accumulated additions and improvements to CS and CSS that have been made since
S16, and that are also reflected in CSDB.  We describe these in order of decreasing substance.  

\subsection{Atomic line list}

The two codes that provide comprehensive spectrum synthesis, CSS and CSDB, now have atomic line lists that are expanded 
as compared to that of CSS as of S16.  As before, all line list data are taken from the NIST Atomic Spectra Database 
\citep{nist}.  The list has been expanded in wavelength coverage to the range 260 to 2600 nm
(previously 360 to 900 nm), and now allows for projects that involve, for example, the important \ion{Mg}{2} hk lines at 280 nm,
and lines in the Johnson {\it H}, {\it J}, and {\it K} photometric bands, which are becoming increasingly important
for chemical abundance analysis of metal-poor red giant stars (see, for example, \citet{rgbir}).  The list has been expanded to include
all ionization stages of H ($Z=1$) through C ($Z=6$) and the first six ionization stages of all other included elements.
The elements represented in the line list now include all those up to Zn ($Z=30$), and heavier elements that have become
increasingly important in the spectroscopic study of old metal poor MWG halo red giants stars, namely Ge ($Z=32$), 
Rb ($Z=37$), Sr ($Z=38$), Y ($Z=39$), Zr ($Z=40$), Cs ($Z=55$), Ba ($Z=56$) and La ($Z=57$) (see, for example, \citet{rgbheavy}, for example).  
The list has also been
expanded to include the weakest transitions - we now include permitted and forbidden transitions with $\log f \ge -6.0$.
The result is that our atomic line list has expanded from $\sim 10000$ to $\sim 26000$ lines.  Again, we emphasize that
both CSS and CSDB have lists that are identical in content - only the implementation differs (byte-data file I/O 
{\it vs} SQL database table).

\subsection{Ionization equilibrium and $n_{\rm e}$ \label{sNe}}

  As of S16, for both CS and CSS, starting from an initial guess, the initial $n_{\rm e}(\tau)$ structure was computed along with the 
initial values of all
 ionization fractions
of all elements and ionization stages being treated, including the important $e^-$ donors in late-type stars (where \ion{H}{1} is neutral), and 
these values were iterated a few times in an attempt to achieve some self-consistency.  We since found that this procedure
is unstable for significant regions of parameter space.  We now pre-compute the $n_{\rm e}(\tau)$ structure from the initial guess
by iteratively solving the more robust implicit expression for the partial $e^-$ pressure, $P_{\rm e}(\tau)$ in terms of the initial 
guess value of the gas pressure, $P_{\rm g}(\tau)$, and the input abundances, $A_{\rm Z}$, assuming all free $e^-$ particles
are contributed by single ionizations only, as described in detail by \citet{gray}.  
This procedure is more reliably stable, and we then proceed
to refine the resulting $n_{\rm e}(\tau)$ structure by iterating it along with the general ionization equilibrium.

 
\subsection{Initial guess structure \label{stemplate}}

As of S16, CS, CSS, and CSDB were equipped with the $T_{\rm kin}(\tau)$, $P_{\rm gas}(\tau)$, and $P_{\rm e}(\tau)$ structures of 
 two research-grade template atmospheric models computed with V. 15 of Phoenix (see Section \ref{srev}):
A representative late-type model of $T_{\rm eff} = 5000$ K and $\log g = 4.5$, and a representative early-type model of
$T_{\rm eff} = 10000$ K and $\log g = 4.0$, both with $[Fe/H] = 0.0$, covering both sides of the major divide between
stars for which the $T_{\rm kin}(\tau)$ structure is, or is not, affected by convection and molecular opacity.  
The details of how these template models are initially re-scaled with the input parameters before structure iterations 
are begun are described in S16.  The $T_{\rm kin}(\tau)$ structure of a dwarf star is
known to over-estimate the $T_{\rm kin}(\tau)$ distribution in the outer atmospheres of lower $\log g$ stars of 
the same $T_{\rm eff}$ value (see, for example, \citet{gray}), and this is especially problematic here given that 
CS, CSS, and CSDB do not currently perform temperature corrections.  This 
leads to an under-estimation of the strength of absorption lines, which predominantly form in the affected layers. 
To improve the $T_{\rm kin}(\tau)$ structure for red giants, we have added 
a third template model computed with Phoenix for $T_{\rm eff} = 4250$ K and $\log g = 2.0$.   If the input parameters
satisfy the condition that $T_{\rm eff} < 7300$ K (MK class F0) and $\log g < 3.5$ then the initial guess is re-scaled
from this template red giant model.

\subsection{Control of execution time {\it vs} realism \label{srealism}}

  As described in Section \ref{srev}, we perform outer iterations among (HSE) ($P_{\rm g}(\tau)$ structure), 
the EOS along with the values of the ionization fractions as coupled by the value of the $n_{\rm e}(\tau)$ structure,
the $\mu(\tau)$ structure, and the $\kappa_\lambda(\tau)$ distribution.  Each outer iteration incorporates an
inner iteration among the ionization fractions 
and the $n_{\rm e}(\tau)$ structure.  The number of both the outer and inner 
iterations is eight by default, and has been found to give approximate convergence for most regions of parameter space at a 
level appropriate for a pedagogical application.  Users who wish to trade off execution time for modeling accuracy, or {\it vice versa},
can separately adjust the number of both the outer and inner iterations in the range from five to 12.  We note for 
comparison that the Phoenix 
modeling code often converges to a few percent accuracy in the energy conservation within ten iterations, even though it is 
also incorporating the thermal equilibrium problem (which CS, CSS, and CSDB are not - see S16), even with the opacity and 
radiation field treated in non-LTE for many cases.  We set a lower limit of five iterations because with fewer, the 
number density of TiO is grossly unrealistic in the coolest models.  The iteration controls are available in the optional
``Performance/realism'' panel of the CS, CSS, and CSDB UIs.    
    
\subsection{Planetary habitable zone \label{swater}}

The number of controls for planetary surface conditions, needed for the computation of the steam and ice lines 
bounding the liquid-surface-solvent habitable zone (HZ), has been expanded from two to three with the addition of the 
atmospheric surface pressure, $P_{\rm surf}$ (the other two are atmospheric greenhouse effect, $\Delta T$, and albedo, {\it A}
 - see S16).  Additionally, the user may now choose from among four simple volatile compounds that were common in parts
of the solar nebula disk: water (H$_{\rm 2}$O), methane (H$_{\rm 4}$C), ammonia (H$_{\rm 3}$N), or carbon dioxide (CO$_{\rm 2}$),
as the liquid solvent for life-bearing organic chemistry.    
The approximate boiling temperature is computed from the input $P_{\rm surf}$ value with the Antoine equation,
 $T_{\rm boil}\approx B/(A-\log P_{\rm surf})-C$, with the
Antoine coefficients, $A$, $B$, and $C$, for $T_{\rm boil}$ in K and $P_{\rm surf}$ in bar, taken from 
the NIST Chemistry WebBook \citep{antoine}.  

The choice of solvent and its boiling point are now displayed on the HZ diagram (along with the
locations of the steam and ice lines, as before).  (We note that the freezing point is negligibly affected by
vapor pressure within the pressure limits that we might expect even microbial life to withstand.)  The addition of a third
planetary parameter and choice of solvent 
enables more complex pedagogical activities in planetary and environmental science - students can 
readily see how the width of the HZ depends on the value of $P_{\rm surf}$, and can, for example, 
investigate how the range of atmospheric greenhouse effect needed to keep a planet at a certain distance
from its host star within the HZ depends on $P_{\rm surf}$. 
The expanded number, and range, of the planetary parameters and choice of organic solvents allows for more
pre-set worlds to be included in the ``Sample worlds'' (formerly ``Sample planets'') section of the optional 
``Samples'' input panel,
and the list now includes Venus, Mars, and the Saturnian moon Titan, as well as Earth. 

\subsection{UI improvements}

  There have been a number of small, but useful, additions to the CS and CSS UIs that are also 
reflected in the CSDB UI.  The reader may refer to S16 for further context: 

\begin{enumerate}

\item{ The wavelength of the narrow-band Gaussian filter through which the narrow band image is made is now indicated on the 
rendering of the classification spectrum.  This is important because students can use the new indicator 
as a guide if they wish to try tuning the filter to
the wavelength of a particular spectral feature - an important experiment that provide a 
starting point for enquiry into how the
$\kappa_\lambda$ distribution affects both the $\cos\theta$ and $\lambda$ variation of the $I_\lambda(\tau=0, \cos\theta, \phi)$
distribution.}

\item{A line indicating a reference distance of 1 AU is now included in the HZ display (see Section 
\ref{swater}).  This provides an important reference because the HZ display is automatically re-scaled to
accommodate the size of the HZ for any given case. }

\item{The HR diagram (HRD) is now marked with lines of constant radii for $R$ values from 0.01 to
1000 R$_{\rm Sun}$ in increments of $\Delta\log_{\rm 10} R = 1.0$ for further context.  This is especially
useful when students are learning to interpret the HRD in terms of the $L = R^2T_{\rm eff}^4$ and
$g = M R^{-2}$ (solar units) relations.}

\item{The background color of the white light image, the HRD, and the HZ displays now automatically
becomes darker for early-type stars that are white or blue-white in color and would formerly lose contrast
with the lighter default background color.} 

\end{enumerate}

\section{Discussion \label{sdiscuss}}

  The stellar astrophysical modeling and visualization WWW applications CS and CSS, and now CSDB, have proven 
to be open-ended projects that can
sustain continual on-going development in the realism and completeness of the physical modeling,
the utility of the UI, and the exploitation of other modern computational tools.  We continue to encourage 
astronomy and physics educators at both the high
school and university level to obtain, and periodically refresh, their own local installation of 
CS and to consider how a responsive model star instrumented in various ways can serve as an apparatus for
classroom demonstration and lab-style homework assignments that help students untangle the many 
physical phenomena that determine stellar observables.  

\paragraph{}

  We encourage instructors of more advanced courses to consider that CSS and CSDB are now 
realistic enough that research projects that are carefully limited in scope, such as a 
differential study of an observed feature over a limited range of stellar parameters, 
is now feasible.  To this end, we remind readers that the modeling output can be optionally printed
as a table, and selected and captured for further processing and analysis in another
environment such as a python IDE. 

\paragraph{}

Based on the feasibility of what has been demonstrated to date, it is clear that the physical modeling
can be further developed while maintaining an execution time appropriate for a pedagogical apparatus.
Additionally, the visualization component can always be further developed by imagining novel and 
insightful ways to post-process and present the large quantity of modeling data produced by a 
model atmosphere and spectrum synthesis code.  We encourage instructors who teach at a more 
advanced level to consider student contributions to such development as an appropriate course
activity.  It is worth noting again here that the basic syntax of the JS and Java programing languages
is identical to that of the scientific programming language C.



\acknowledgments
The author acknowledges Natural Sciences and Engineering Research Council of 
Canada (NSERC) grant RGPIN-2014-03979.  We also thank David Lane for help
with the Java-MySql interaction, and IT professional
Byron Clairoux for suggesting the potential suitability of 
SQL databases for direct spectrum synthesis.  We thank Charli Sakari for important 
feedback from high school and college teachers in the University of Washington in the 
High Schools, and in the Northwest Astronomy Teaching Exchange, programs. 
The author thanks the astronomy faculty, staff, and students at the University of British Colombia,
the University of Victoria, and the University of Washington for helpful 
suggestions following demonstrations of CS at their campuses in the winter of 2016.

\clearpage



\clearpage






\end{document}